# Identifying intrinsic and extrinsic mechanisms of anisotropic magnetoresistance with terahertz probes

Ji-Ho Park[1], Hye-Won Ko[1], Jeong-Mok Kim[2], Jungmin Park[3], Seung-Young Park[3], Younghun Jo[3], Byong-Guk Park[2], Se Kwon Kim[1], Kyung-Jin Lee[1,4,5], and Kab-Jin Kim[1]★

[1]*Department of Physics, KAIST, Daejeon 34141, South Korea*

[2]*Department of Materials Science and Engineering and KI for Nanocentury, KAIST, Daejeon 34141, South Korea*

[3]*Center for Scientific Instrumentation, KBSI, Daejeon 34133, South Korea*

[4]*Department of Materials Science & Engineering, Korea University, Seoul 02841, South Korea*

[5]*KU-KIST Graduate School of Converging Science and Technology, Korea University, Seoul 02841, South Korea*

★ Correspondence to: kabjin@kaist.ac.kr

**Identifying the intrinsic and extrinsic origins of magneto-transport in spin-orbit coupled systems has long been a central theme in condensed matter physics. However, it has been elusive owing to the lack of an appropriate experimental tool. In this work, using terahertz time-domain spectroscopy, we unambiguously disentangle the intrinsic and extrinsic contributions to the anisotropic magnetoresistance (AMR) of a permalloy film. We find that the scattering-independent intrinsic contribution to AMR is sizable and is as large as the scattering-dependent extrinsic contribution to AMR. Moreover, the portion of intrinsic contribution to total AMR increases with increasing temperature due to the reduction of extrinsic contribution. Further investigation reveals that the reduction of extrinsic contribution is caused by the phonon/magnon-induced negative AMR. Our result will stimulate further researches on other spin-orbit-interaction-induced phenomena for which identifying the intrinsic and extrinsic contributions is important.**

Understanding charge and spin transport in spin-orbit-coupled systems is a central theme of condensed matter physics. Representative examples are anomalous Hall effect (AHE)[1] and spin Hall effect (SHE)[2]. They share the microscopic origins: The intrinsic scattering-independent mechanism due to Berry phase effects of the electron band structure[3] and the extrinsic scattering-dependent mechanism due to skew scattering[4] and side jump[5] have been identified. The fundamental understanding of the AHE and SHE has been important milestones in condensed matter physics[1,2] and has recently led to practical applications based on the spin-orbit interaction (SOI), e.g. spin-orbit torque (SOT) devices[6-8], where identifying intrinsic or extrinsic mechanism[9,10] is crucial for efficient device operation.

The AMR[11] is a fundamental SOI-induced transport phenomenon in magnetic materials. It describes the anisotropic charge conductivity depending on relative orientations of the current flow and the magnetization. Early theories viewed the AMR through the Mott's two-current model[12], in which the electrical conduction in transition metals is modelled as the sum of two separate currents of majority and minority spin electrons with following two assumptions: (i) $s$ electrons are responsible for the conduction owing to relatively low mobility of $d$ electrons, (ii) large density of $d$ states mainly accounts for scattering rates of $s$ electrons. Adopting this point of view resulted in the extrinsic magnetization-dependent scattering time as an origin of AMR[13-17], since the SOI works for $d$ electrons but not for $s$ electrons. In other words, conducting $s$ electrons experience anisotropic scattering processes via spin-orbit-coupled $d$ bands. This class of AMR theories considered various SOI-induced effects on the scattering time such as a mixing of majority and minority $d$ states[15-17], an alteration of atomic wave function of $d$ orbitals[16], or a spin mixing process including spin-flip scattering[14].

Recent theories have relaxed the assumptions of Mott model and identified the intrinsic mechanism of AMR, which arises from the scattering-independent band-structure effect. This relaxation is essential for transition metals since $d$ electrons also participate in electrical conduction[18,19] through $spd$ hybridization[20]. In other words, the conduction electrons are no longer intact to SOI; rather, their band structures are affected by SOI. Therefore, not only the magnetization-dependent scattering time but also the magnetization-dependent electronic structure, i.e., the intrinsic origin, plays a crucial role in the AMR. Recent theoretical works have emphasized the intrinsic effect on AMR by showing a ballistic AMR owing to the anisotropic $d$ bands[21] and the magnetization-dependent band structure owing to the orbital hybridization[22]. The existence of the intrinsic contribution has also been addressed experimentally. Hupfauer *et al.* showed the crystalline AMR effect originating from the change of electronic density by magnetization orientation[23], and Zeng *et al.* reported the intrinsic contribution caused by the magnetization-direction-dependent band crossing effect[24].

Despite the advances in the fundamental origin of AMR, however, our understanding is far from complete. In particular, the quantitative separation between intrinsic and extrinsic contributions to AMR in real systems still remains elusive. We note that the transverse AMR (also known as the planar Hall effect), which is concomitant with the longitudinal AMR, generates a SOT[25-27]. In this respect, identifying the intrinsic and extrinsic AMR mechanisms is of crucial importance not only for fundamental understanding of SOI-induced transport in magnetic materials but also for practical application based on the AMR, as for the AHE and SHE.

As the AMR stems from the change in the longitudinal conductivity, one can use the DC Drude model[28] to investigate the intrinsic and extrinsic contributions to the AMR:

$$\sigma_{dc} = \frac{n}{m^*} e^2 \tau, \qquad (1)$$

where $\sigma_{dc}$ is the DC conductivity, $n$ is the charge density, $m^*$ is the effective mass, $e$ is the electron charge, and $\tau$ is the momentum scattering time. Here the magnetization-dependent change in $n/m^*$ represents the intrinsic scattering-independent contribution whereas the magnetization-dependent change in $\tau$ represents the extrinsic scattering-dependent contribution. Therefore, the intrinsic and extrinsic contributions to the AMR can be separately identified by measuring $\frac{n}{m^*}$ and $\tau$ for different magnetization directions. In the present work, we achieve this separate identification by measuring AC Drude conductivity with the terahertz (THz) time-domain spectroscopy (TDS), which was applied to understand the giant magnetoresistance in a previous work[29].

The AC Drude conductivity is given by

$$\tilde{\sigma}(\omega) = \frac{\sigma_{dc}}{1 - i\omega\tau} = \frac{\sigma_{dc}}{1 + \omega^2\tau^2} + i\omega\tau \frac{\sigma_{dc}}{1 + \omega^2\tau^2}, \qquad (2)$$

where $\omega/2\pi$ is the frequency. THz spectroscopy allows us to determine the real and imaginary components of the AC Drude conductivity, leading to the direct determination of the momentum scattering time $\tau$ and DC conductivity $\sigma_{dc}$, simultaneously, based on Eq. (2). Therefore, it enables the separate quantification of $\tau$ and $n/m^*$ using Eq. (1), which correspond to the extrinsic and intrinsic contributions to AMR, respectively.

Figure 1a shows the schematic illustration of our THz-TDS setup. The 90-nm-thick permalloy (Py) film, deposited on Si/SiO$_2$ substrate by magnetron sputtering, is located inside the cryostat. The weak single-cycle, sub-picosecond THz pulse is directed normally at the sample (along *z*-axis), and its transmission is measured. The polarization of THz electric field lies along the *y*-axis, and the magnetization direction of Py is controlled in the *x-y* plane by applying in-plane magnetic field using home-built vector magnet (see Supplementary Notes 1-2 for sample and setup details).

Figure 1b shows the transmitted THz time-domain signal by changing the angle between the THz electric field and the Py magnetization. Here we applied magnetic field of 25 mT, which is large enough to saturate the magnetization of Py film. The measurement was conducted at *T* = 54 K. To increase the signal to noise ratio, we averaged the time-domain trace by accumulating 850 pulses for each measurement, and this measurement was repeated 40 times (see Supplementary Notes 3 for measurement details). As the oscillating THz electric field induces a time-dependent current in the sample, the attenuation and phase delay of THz electric field can be observed as it propagates through the sample. Since the THz frequency is lower than the electron relaxation rate, the degree of attenuation depends on the DC resistivity of sample: the higher the resistivity, the smaller the attenuation. Since the Py has a positive AMR[30], that is $\rho^{AMR} \equiv \frac{\rho^{\parallel}-\rho^{\perp}}{\rho^{\perp}} > 0$, where $\rho^{\parallel}(\rho^{\perp})$ is the resistivity of Py when the current and magnetization are parallel (perpendicular), the resistivity of sample is high (low) when the magnetization and electric current is parallel (perpendicular). Therefore, a large (small) attenuation of THz electric field is expected when the THz electric field and Py magnetization is perpendicular (parallel) to each other. This is indeed observed in our THz time domain signals in Fig. 1b.

To extract the intrinsic and extrinsic parameters, we obtained complex conductivity of sample by performing Fourier transform from the THz time domain signals (see Supplementary Note 3 for detailed process). Figure 1c shows the THz spectra of real and imaginary parts of complex conductivity for 5 different angles (90°, 40°, 0°, -40°, -90°). Solid lines are the best fits by the AC Drude model based on Eq. (2). The coincidence of experimental data and AC Drude fitting demonstrates that the complex conductivity of Py can be well described by the AC Drude model. The real and imaginary components of complex conductivities directly yield the momentum scattering time $\tau$ and the DC resistivity $\left(\rho_{dc,THz} = \frac{1}{\sigma_{dc,THz}}\right)$ at the same time, when using Eq. (2). Here, $\rho_{dc,THz}$ denotes the DC conductivity obtained by THz-TDS. The validity of $\sigma_{dc,THz}$ was confirmed by separate four-probe DC resistance measurement (see Supplementary Note 4). We plot those values in Fig. 1d. Both the $\tau$ (blue symbols) and the $\rho_{dc,THz}$ (black symbols) follow the typical AMR angle dependence (~$\cos^2\theta$ shown in solid lines in Fig. 1d), and they are inversely proportional to each other, in accordance with $\frac{1}{\rho_{dc,THz}} = \sigma_{dc,THz} = \frac{n}{m^*}e^2\tau$. However, we found that the amount of anisotropy is different for $\rho_{dc,THz}$ (4.3 %) and for $\tau$ (2.8 %), which implies that not only extrinsic $\tau$ but also the intrinsic $n/m^*$ contributes the anisotropy of $\rho_{dc,THz}$.

To further investigate the intrinsic and extrinsic contribution to AMR, we perform the temperature dependence of intrinsic and extrinsic parameters by repeating the experiment at various temperatures. To enhance the accuracy, we performed THz-TDS measurement for parallel and perpendicular geometries, repeatedly up to 160 times at a fixed temperature. The measurement temperature was varied in the range of $T <$ 155 K where the Drude fitting can be validated; at higher temperatures the uncertainty in determining $\tau$ increases due to the rapid decrease of $\tau$ and the small AMR, so the error bars of $\tau$ for parallel and perpendicular geometries are overlapped with each other. Figures 2a-2c show the temperature dependence of $\rho_{dc,THz}$ [Fig. 2a], $\tau$ [Fig. 2b], and $n/m^*$ [Fig. 2c] for parallel (black) and perpendicular (red) geometries. Here $\rho_{dc,THz}$ and $\tau$ are directly determined from the THz-TDS measurement, and $n/m^*$ is extracted from the measured $\rho_{dc,THz}$ and $\tau$ based on Eq. (1). Figures 2(a) and 2(b) show that, as we increase the temperature, the $\rho_{dc,THz}$ increases but the $\tau$ decreases, which are typical trends of metallic samples originating from the increasing effect of electron scattering by thermally excited phonons and magnons[31]. On the other hand, Fig. 2c shows that not only the scattering-dependent $\tau$, but also the scattering-independent $n/m^*$ also changes in temperature. To be more quantitative, the $\tau$ decreases by about 20%, while the $n/m^*$ decreases by about 7% with increasing temperature from $T$ = 16 K to $T$ = 155 K. Since the $n/m^*$ originates from the intrinsic band structure, the reduction of $n/m^*$ might be caused by, for examples, expansion of lattice[32], temperature dependent exchange constant[33-35] and electron mass enhancement[36, 37] (see Supplementary Note 5 for the detailed discussion on the temperature dependence of $n$ and $m^*$).

The clear difference between parallel and perpendicular geometries in Figs. 2a-2c indicates that all parameters are anisotropic, and thus contribute to the AMR. In Fig. 3, we summarize the temperature dependent anisotropy for $\rho^{AMR} \equiv \frac{\rho_{dc,THz}^{\parallel} - \rho_{dc,THz}^{\perp}}{\rho_{dc,THz}^{\perp}}$ [gray] and $\tau^{AMR} \equiv \frac{\tau^{\perp} - \tau^{\parallel}}{\tau^{\parallel}}$ [green], as well as for $(n/m^*)^{AMR} \equiv \frac{(n/m^*)^{\perp} - (n/m^*)^{\parallel}}{(n/m^*)^{\parallel}}$ [blue], which are extracted from Figs. 2a-2c. The result suggests that the total AMR ($\rho^{AMR}$) consists of not only the extrinsic contribution induced by electron scattering ($\tau^{AMR}$) but also the intrinsic contribution induced by band-structure [$(n/m^*)^{AMR}$], as can be written by $\rho^{AMR} = \tau^{AMR} + (n/m^*)^{AMR}$ in the first order approximation. We note that the band-induced intrinsic contribution can exist even in polycrystalline samples, like the intrinsic anomalous Hall effect observed in polycrystalline samples[38] (for further detailed discussion, see Supplementary Note 6).

The temperature dependence of extrinsic and intrinsic contribution shows distinct behaviour: the $\tau^{AMR}$ largely decreases with increasing temperature even considering error bars [green symbols in Fig. 3], while the $(n/m^*)^{AMR}$ does not exhibit a clear change but is still in the range of error bars [blue symbols in Fig. 3]. To be more quantitative, the $\tau^{AMR}$ drops by about 50% as the temperature increases from $T$ = 16 K to $T$ = 155K, while $n/m^{*AMR}$ remains almost constant within $\pm$22% error range. This means that the portion of intrinsic contribution to total AMR gradually increases with increasing temperature, mainly due to the reduction of extrinsic contribution at higher temperature. The intrinsic portion is 32.3 $\pm$ 7.3% of total AMR at $T$ = 16 K, but it increases 46.2 $\pm$ 9.7 % at $T$ = 155 K. The large intrinsic contribution, which is comparable to the extrinsic contribution, suggests that the intrinsic contribution is important at room temperature, e.g., for SOT devices utilizing the transverse AMR as a spin-current source[25-27].

The large reduction of extrinsic contribution at higher temperature is further investigated. The scattering-induced extrinsic contribution can be categorized into two sources: electron-impurity scattering and electron-phonon/magnon scattering[39]. Considering the Matthiessen's rule, the resistivity and scattering time can be expressed by

$$\rho = \rho_{e-im} + \rho_{e-ph/mag}, \qquad (3)$$

$$\frac{1}{\tau} = \frac{1}{\tau_{e-im}} + \frac{1}{\tau_{e-ph/mag}}, \qquad (4)$$

where, $\rho_{e-im}$ ($\tau_{e-im}$) and $\rho_{e-ph/mag}$ ($\tau_{e-ph/mag}$) represent the DC resistivities (scattering times) induced by electron-impurity scattering and electron-phonon/magnon scattering, respectively. As phonons and magnons are thermally excited quasiparticles, their effect becomes negligible at low temperature limit[31, 40]. Therefore, the THz-probed DC resistivity and scattering time measured at 16 K (the lowest temperature in our THz setup) can be considered as results from pure electron-impurity scattering (see Supplementary Fig. S4 for $\rho - T$ curve at low temperature). As the electron-impurity scattering does not depend on the temperature, we can separate the contributions from impurity and phonon/magnon based on Eqs. (3) and (4). Figures 4a and 4b show the temperature dependent $\rho_{e-ph/mag}$ and $\tau_{e-ph/mag}$ for parallel (black) and perpendicular (red) geometries. The $\rho_{e-ph/mag}$ increases rapidly with increasing temperature, while the $\tau_{e-ph/mag}$ decreases drastically with temperature, indicating the increasing effect of thermal excitation at higher temperature. The inset of Fig. 4b shows the electron-phonon/magnon scattering rate ($1/\tau_{e-ph/mag}$). The solid lines in Fig. 4a and inset of Fig. 4b show the fitting lines based on $\rho_{e-ph/mag} = aT^5$ and $1/\tau_{e-ph/mag} = bT^3$, respectively[41]. This specific temperature dependence suggests that the extracted resistivities and scattering times indeed arise from the electron-phonon/magnon scattering.

The phonon/magnon-induced AMR, i.e., $\rho_{e-ph/mag}^{AMR}(T) = \left|\frac{\rho_{e-ph/mag}^{\parallel}(T) - \rho_{e-ph/mag}^{\perp}(T)}{\rho_{e-ph/mag}^{\parallel}(T)}\right|$, is extracted from Fig. 4a, and plotted as orange symbols in Fig. 4c. For comparison, we denoted the impurity-induced AMR obtained at $T$ = 16 K, i.e., $\rho_{e-im}^{AMR} = \left|\frac{\rho_{dc}^{\parallel}(T=16K) - \rho_{dc}^{\perp}(T=16K)}{\rho_{dc}^{\parallel}(T=16K)}\right|$ as a purple

dashed line in Fig. 4c. The results clearly show that the sign of phonon/magnon-induced AMR is negative, while that of the impurity-induced AMR is positive. The negative phonon/magnon-induced AMR is also confirmed by the temperature dependent standard DC resistivity measurement, which is shown as black symbols in Fig. 4c (data are extracted from Fig. S4). This indicates that the negative phonon/magnon-induced AMR can be an origin of the reduction of the extrinsic AMR at higher temperature. We note that the negative AMR induced by phonon and magnon possibly originates from the lifting of degeneracy in virtual bound states due to the $L_Z S_Z$ part of the spin-orbit coupling[42,43]. However, more rigorous theoretical studies are required to understand the exact origin of temperature dependence of phonon/magnon-induced AMR, which is beyond the scope of this work.

To conclude, we have unambiguously distinguished the intrinsic and extrinsic contributions to the AMR by independently probing the electron momentum scattering time ($\tau$) and charge density/effective mass ($n/m^*$) using THz-TDS. We found that the AMR originates not only from the scattering-dependent extrinsic mechanism via $\tau$ but also from the scattering-independent intrinsic mechanism via $n/m^*$, revealing the multiple channels through which the SOI gives rise to the AMR. The intrinsic contribution was found to be comparable to the extrinsic contribution, indicating the importance of the band anisotropy in AMR that has been overlooked so far. Our results therefore provide the experimental evidences for the distinct microscopic mechanisms underlying the AMR, and call for further theoretical investigation for complete understanding of AMR with considering the intrinsic and extrinsic mechanisms on an equal footing. Finally, we anticipate that our work will stimulate fundamental studies for separating intrinsic and extrinsic origins of various other spin-transport phenomena in magnetic materials, such as the anomalous Hall effect and magnetic damping[44,45].

*Note added*: After submission of this manuscript, we became aware that L. Nadvornik et al. reported a similar finding[46]. Their approach is based on the wide-frequency Drude fitting from diffusive to ballistic transport regime, while our approach is based on direct determination of $\tau^{AMR}$ (($n/m^*)^{AMR}$) by measuring $\tau^\parallel$ and $\tau^\perp$ ( $(n/m^*)^\parallel$ and $(n/m^*)^\perp$ ). The different results between two groups may be due to the different film thickness, which could result in interface-dominant transport at thin layer or bulk-dominant transport at thick layer[47].

**Acknowledgements**

We would like to thank anonymous referees for valuable comments. This work was supported by the National Research Foundation of Korea (NRF) (Grant Nos. NRF-2016R1A5A1008184, 2020R1A2C4001789). S.K.K was supported by Brain Pool Plus Program through the National Research

Foundation of Korea funded by the Ministry of Science and ICT (Grant No. NRF-2020H1D3A2A03099291). K.-J.L. was supported by the National Research Foundation (NRF) of Korea (Grant No. NRF- 2020R1A2C3013302). B.-G.P. was supported by the National Research Foundation (NRF) of Korea (Grant No. NRF- 2020R1A2C2010309). J. P, S.-Y. P. Y. J acknowledge the support by the National Research Council of Science and Technology (NST) (Grant No. CAP-16-01-KIST) by the Korea government (MSIP).14

**Figure Legends**

**Figure 1| Angle-dependent complex conductivity measurement. a**, Schematic illustration of terahertz time-domain spectroscopy (THz-TDS). **b**, Time-domain THz pulses for several angles between THz electric field and magnetization of sample. Inset zooms into the amplitude of transmitted pulses. The measurement temperature is $T = 54$ K. **c**, The real and imaginary parts of complex conductivity with respect to the THz frequency for several angles. Solid lines are the Drude fittings using Eq. (2). **d**, Angle-dependent DC resistivity (black) and scattering time (blue) determined by THz spectroscopy.

**Figure 2| Temperature-dependent intrinsic and extrinsic transport parameters probed by THz-TDS. a**, THz-proved DC resistivity, $\rho_{dc,THz}$, as a function of temperature for parallel (black) and perpendicular (red) geometries. **b**, Temperature-dependent electron scattering time, $\tau$, for parallel (black) and perpendicular (red) geometries. **c**, Temperature-dependent charge density/effective mass, $n/m^*$, for parallel (black) and perpendicular (red) geometries.

**Figure 3| Intrinsic and extrinsic contributions to total AMR.** Anisotropy ratios of $\rho_{dc,THz}$ (gray), $\tau$ (green) and $(n/m^*)$ (blue) as a function of the temperature. Here, $\rho^{AMR}$ corresponds to the total AMR, while $\tau^{AMR}$ and $(n/m^*)^{AMR}$ represent the extrinsic and intrinsic contribution to AMR.

**Figure 4| Extrinsic AMR induced by phonons and magnons a.** Phonon/magnon-mediated DC resistivity, $\rho_{e-ph/mag}$, as a function of temperature for parallel (black) and perpendicular (red) geometries. Solid lines are the best fits for $\rho_{e-ph/mag} = aT^5$. **b.** Phonon/magnon-mediated electron scattering time, $\tau_{e-ph/mag}$, as a function of temperature for parallel (black) and perpendicular (red) geometries. Inset shows the temperature dependent phonon/magnon-mediated scattering rate, $1/\tau_{e-ph/mag}$. Solid lines in the inset are the best fits for $1/\tau_{e-ph/mag} = bT^3$. **c.** The extracted

phonon/magnon-mediated AMR (orange symbols). The purple dashed line indicates the temperature-independent impurity-mediated AMR. The black symbols denote the phonon/magnon-mediated AMR obtained from the standard DC resistivity measurement shown in Fig. S4.

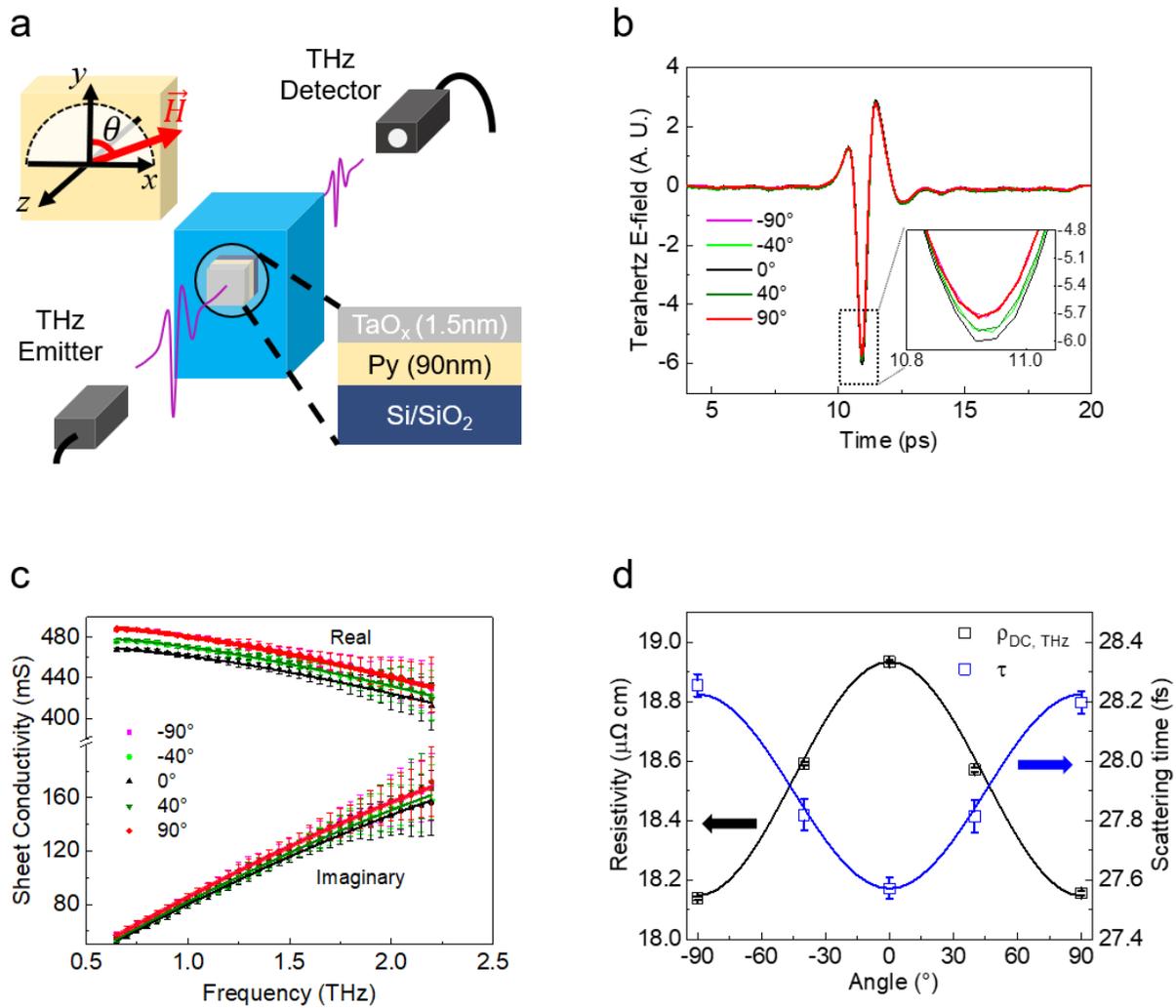

**Fig. 1**

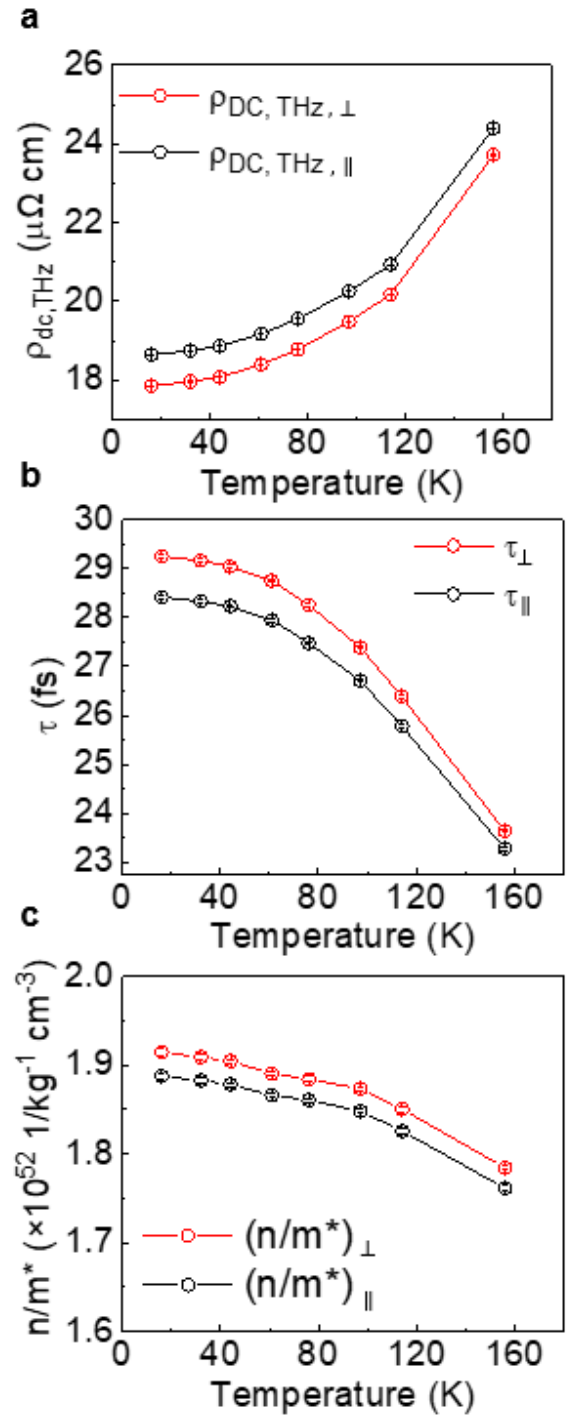

**Fig. 2**

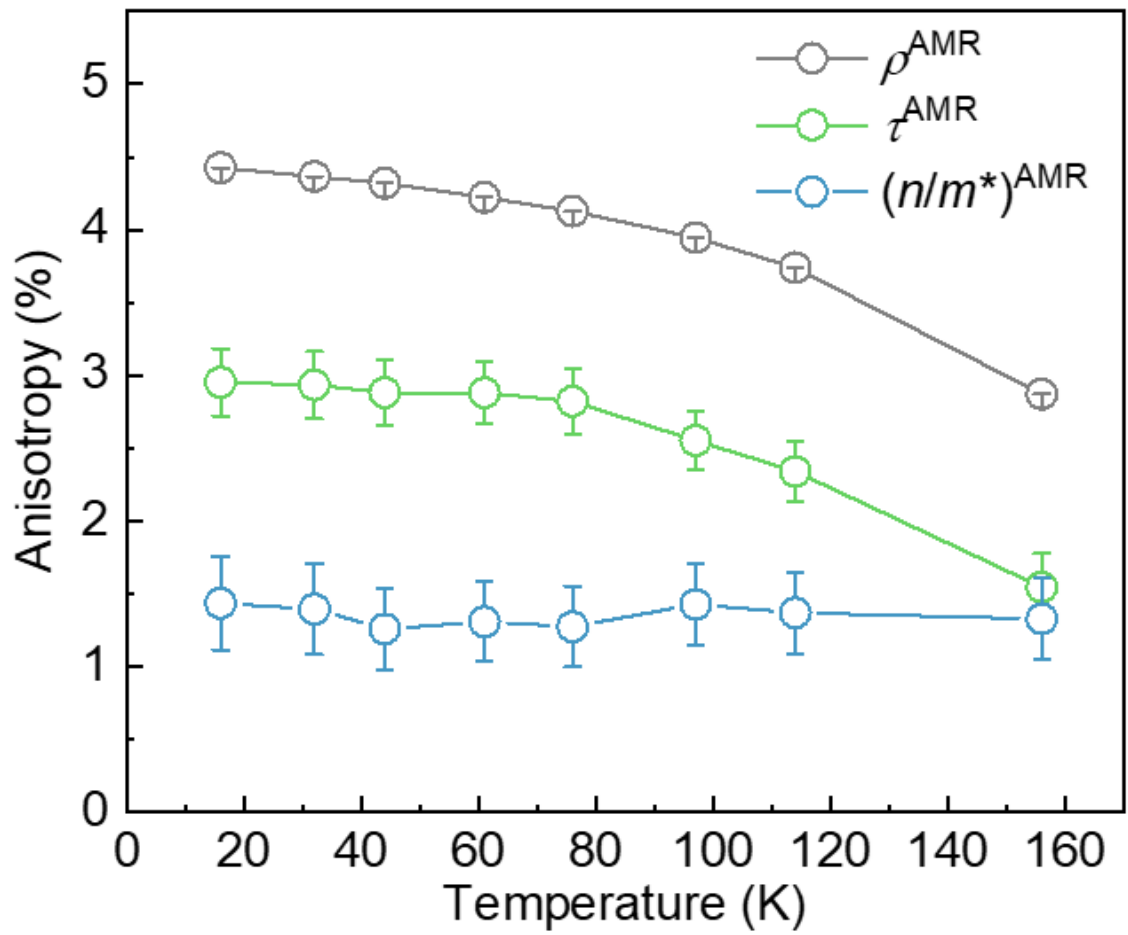

**Fig. 3**

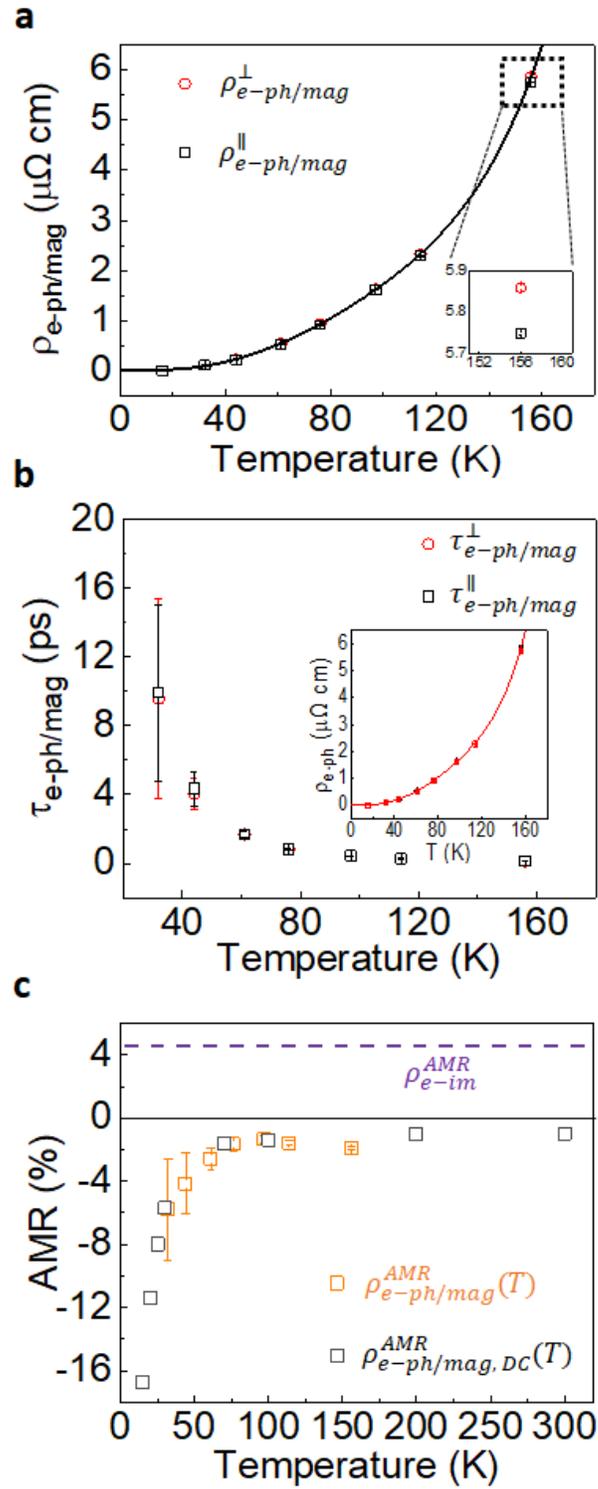

**Fig. 4**

# Supplementary Note for

# "Identifying intrinsic and extrinsic mechanisms of anisotropic magnetoresistance with terahertz probes"

Ji-Ho Park[1], Hye-Won Ko[1], Jeong-Mok Kim[2], Jungmin Park[3], Seung-Young Park[3], Younghun Jo[3], Byong-Guk Park[2], Se Kwon Kim[1], Kyung-Jin Lee[1,4,5], and Kab-Jin Kim[1]★

[1]*Department of Physics, KAIST, Daejeon 34141, South Korea*

[2]*Department of Materials Science and Engineering and KI for Nanocentury, KAIST, Daejeon 34141, South Korea*

[3]*Center for Scientific Instrumentation, KBSI, Daejeon 34133, South Korea*

[4]*Department of Materials Science & Engineering, Korea University, Seoul 02841, South Korea*

[5]*KU-KIST Graduate School of Converging Science and Technology, Korea University, Seoul 02841, South Korea*

## -Contents-

**Note 1. Sample preparation**

90-nm-thick Py films with 1.5-nm $TaO_x$ capping layer were deposited by rf magnetron sputtering on $Si/SiO_2$ substrate. The Py films had in-plane magnetic anisotropy, and the size of films was $6 \times 12$ mm$^2$ which was larger than the diffraction limit of THz wave (3 mm). To measure the complex conductivity of sample, we prepared bare $Si/SiO_2$ substrate as a reference which has almost identical thickness with that of the deposited sample. For this study, we prepared three Py films (#1 - #3): #1 was used for angle dependent THz-TDS measurement (Fig. 1), #2 was used for temperature dependent THz-TDS measurement (Figs. 2,3 and 4), and #3 was used for four-probe DC resistance measurement (inset of Fig. S4). Nominal thickness of samples was same for all samples.

**Note 2. Experimental setup for THz-TDS**

Montana vacuum cryostat was used to control the temperature of sample. A standard glass windows were replaced with the TPX windows for THz experiment. The deposited film and reference substrate were installed simultaneously inside the cryostat using L-shaped sample holder. Two samples (deposited film and reference substrate) were attached to each plane of L-shaped holder that was attached at the ANC300 piezo rotator. By rotating the piezo rotator, we could selectively choose the deposited film or reference substrate without breaking the vacuum. An in-plane magnetic field was applied by using home-made vector electromagnet. The vector magnet is made of two axis electromagnets, each of which generates magnetic field of up to 100 mT. The directional magnetic field can be applied by adjusting the strengths of two orthogonal magnetic fields. The direction of the magnetic field is confirmed by sensing the magnetic field at sample position using Gaussmeter. To rule out any mechanical effects, every structure inside the vacuum chamber is made of OFHC (Oxygen free high thermal conductivity) copper and brass which have very low magnetic susceptibility[1]. A standard THz-TDS setup (Tera K15-Menlosystems) with 4 TPX lenses was used for THz measurement.

To check any mechanical effects from the magnetic field, we prepared non-magnetic Pt (5 nm)/Ta

(1 nm) sample. This sample does not show any AMR but may exhibit other field-induced artifacts if any. Figure S1 shows the transmitted THz pulses for various field orientations (and also for field zero). As shown in the figure, we did not observe any noticeable change by magnetic field in Pt/Ta sample, which implies that the magnetic field does not affect the AMR measurement. This additional experiment can also rule out the possibility of ordinary magnetoresistance effect induced by Lorentz force, which was estimated to be negligibly small (~$4 \times 10^{-8}$ % from $\Delta\rho/\rho = \mu^2 B^2$, where $\mu$ (~$10^{-3} m^2/V \cdot s$) is the mobility and $B$ (= $20 mT$) is the external field ) compared to the AMR change (3~4%).

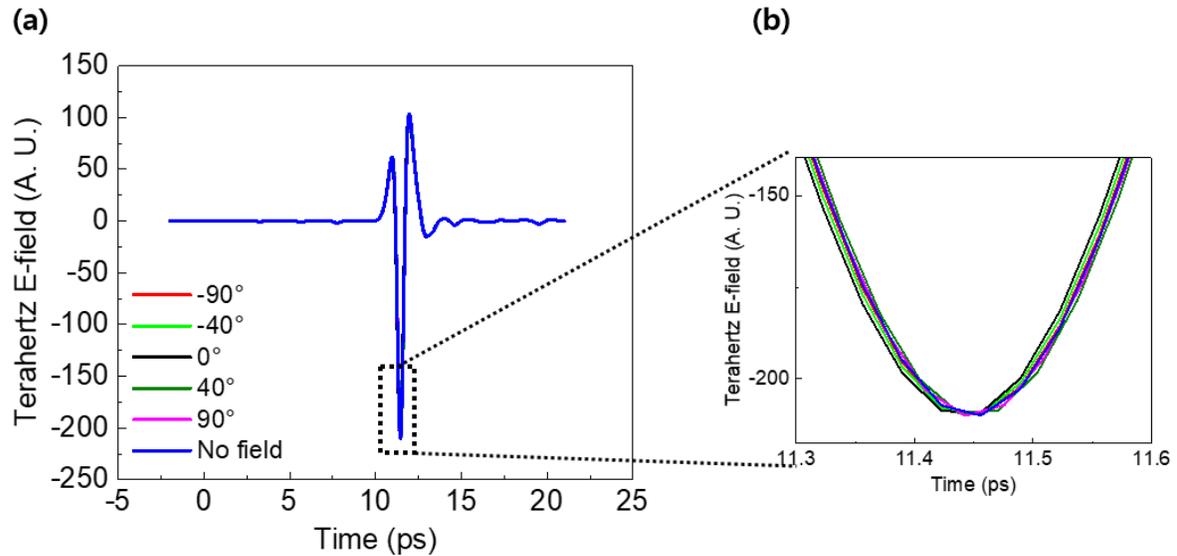

**Fig.S1** (a) Transmitted terahertz pulse depending on the field orientation for Pt (5 nm)/Ta (1 nm) film. Blue curve represents the result without magnetic field. (b) Enlarged pulse peaks.

## Note 3. THz measurement and analysis procedure

**THz pulse detection and Fourier transform**

A single cycle THz pulse with 2ps duration was generated from the THz emitter and then, the THz beam was focused on sample through the TPX lens. The transmitted pulse was collimated and detected at the THz detector. The polarization of THz electric field was fixed along *y*-axis (Fig. 1a). Figure S2(a) shows the THz pulse, typically obtained in our THz-TDS setup, and Fig. S2(b) is enlarged pulse. The symbols are experimental data that are actually obtained from the machine, and the lines are the guide to the eye. The time interval between neighboring symbols is 33 $fs$, which corresponds the time resolution of our THz-TDS setup. This means that the theoretical frequency maximum that we can access is about 15 THz (=1/(2 × 33$fs$)), but in real system the accessible frequency range is reduced due to the noise and large attenuation from the sample. To do the frequency analysis, we perform the Fourier transformation from the time domain pulse data. Fig. S2(c) shows the typical THz spectra obtained from the Fourier transform. As shown in the figure, the meaningful frequency range is limited under 2.5 THz. We note that the THz spectra in our result shows very smooth shape without any dips or peaks. This is important because a dip or peak in spectral amplitude leads to unwanted dip or peak in complex conductivity, which causes a large error in Drude fitting.

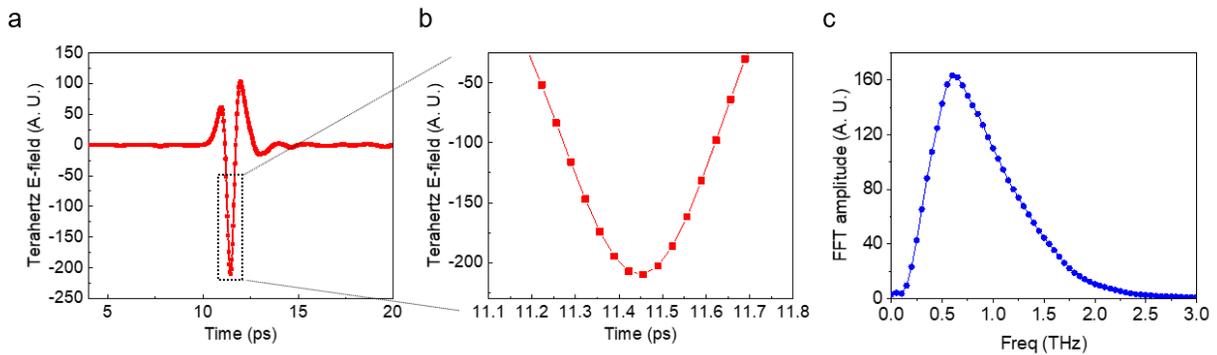

**Fig.S2** (a) Typical time domain THz pulse (b) Enlarged THz pulse. (c) Typical THz spectra obtained by Fourier transform from time domain pulse.

**Complex conductivity measurement**

To obtain complex sheet conductivity, the detected time domain signal was first converted to the frequency domain by Fourier transform, which provides the frequency-dependent amplitude $A(\omega)$ and phase $\phi(\omega)$ of electric field $E(\omega)$. This procedure was repeated for reference substrate, and obtained $A_{\text{ref}}(\omega)$ and $\phi_{\text{ref}}(\omega)$ of $E_{\text{ref}}(\omega)$. The obtained amplitudes and phases were then inserted in the following Tinkham equation[2]

$$\frac{E(\omega)}{E_{\text{ref}}(\omega)} = \frac{n_{sub}+1}{n_{sub}+1+Z_0\tilde{\sigma}_s(\omega)}, \quad (1)$$

where $Z_0$ is the vacuum impedance and $n_{sub}$ is the refractive index of substrate. By inserting the experimentally obtained $E(\omega) = A(\omega)e^{i\phi(\omega)}$ and $E_{\text{ref}}(\omega) = A_{\text{ref}}(\omega)e^{i\phi_{\text{ref}}(\omega)}$ into Eq. (1), the complex conductivity reads

$$\tilde{\sigma}_s(\omega) = \left[\left(\frac{A_{\text{ref}}(\omega)e^{i\phi_{\text{ref}}(\omega)}}{A(\omega)e^{i\phi(\omega)}}\right) - 1\right] * \frac{[n_{sub}+1]}{Z_0}$$

$$= \left[\left\{\frac{A_{\text{ref}}(\omega)}{A(\omega)}\cos(\phi_{\text{ref}}(\omega) - \phi(\omega)) - 1\right\} + i\left\{\frac{A_{\text{ref}}(\omega)}{A(\omega)}\sin(\phi_{\text{ref}}(\omega) - \phi(\omega))\right\}\right] * \frac{[n_{sub}+1]}{Z_0}, \quad (2)$$

Then one can obtain the real and imaginary parts of complex sheet conductivity $\tilde{\sigma}_s$. Here, the refractive index of substrate was obtained by independent experiment[3]. Once the real and imaginary conductivities were obtained, the scattering time and DC conductivity can be directly determined by using Eq. (2) in the main manuscript without any assumptions on charge density and effective mass.

**Noise reduction and average process**

To reduce the noise, we performed the following treatments: 1) THz system was purged with dry nitrogen gas until the relative humidity was down to less than 1%. 2) A 675μm thick substrate was chosen to keep internally reflected pulse away from the main pulse. 3) Time domain was carefully selected in order to exclude the effect of internal reflections which would cause the ripple in the frequency domain. 4) High resistive Si was used to reduce the absorption loss in substrate. 5) Both sides of substrate were polished to reduce the effect from surface roughness which would affects the transmittance of THz pulse. 6) The thickness of Py (90 nm) was chosen to increase the difference of the sheet conductivity without significantly deteriorating the signal-to-noise ratio.

To increase the signal to noise ratio, we repeated the measurement and averaged them. The size of error bars in Fig. 1(c) in our manuscript was obtained from 1σ value of 34,000 data (850 pulses at each measurement × 40 repeated measurements). For the temperature dependent measurement shown in Fig. 2, we further averaged up to 136,000 data (850 pulses at each measurement × 160 repeated measurements) to reduce the size of error bars. We note that the error during the repeated measurement mainly stems from the phase jitter of our THz-TDS system. Due to the noise reduction as discussed above, the resulting jitter was confirmed to be smaller than time resolution (33 *fs*) of our THz setup.

Figure S3 shows the raw data sets for Fig. 2 in the main manuscript, which exhibit the real and imaginary spectra for various temperatures. One can notice that the error bars are further reduced from Fig. 1(c) by averaging process.

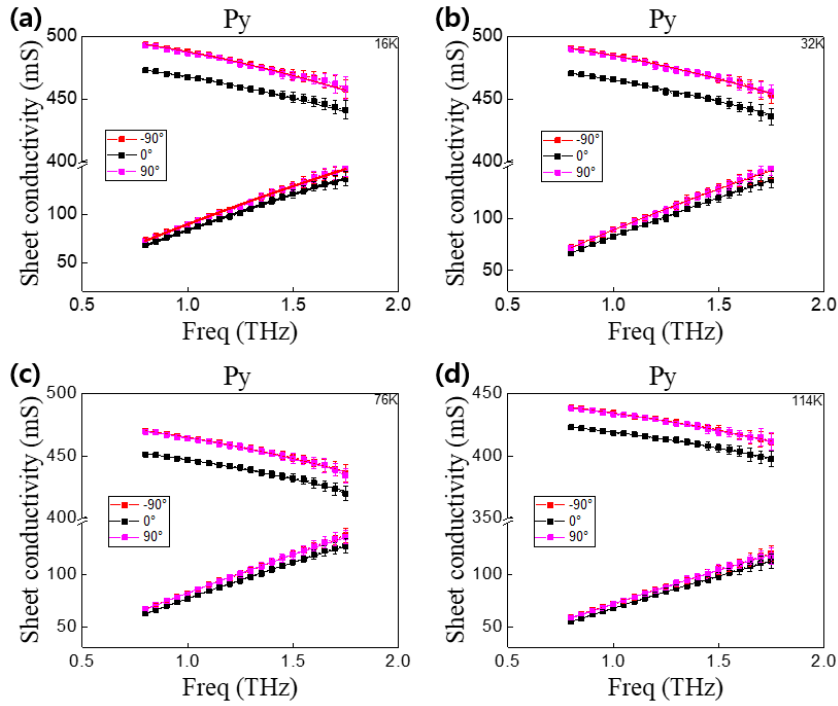

**Fig. S3** The real and imaginary parts of complex conductivity with respect to the THz frequency for several angles. Solid lines are the Drude fittings. (a) 16K (b) 32K (c) 76K (d) 114K.

**<u>Condition for Drude fitting</u>**

To extract the scattering time, we used the AC Drude model. Based on the Drude model, we fitted the real and imaginary part of complex conductivity as shown in Fig. S3 (solid lines are the best Drude fit). Here, we note that, for Drude fitting, the frequency range should satisfy two conditions. First, to apply Drude fitting, the frequency range should be $\omega\tau < 1$ (or $f < \frac{1}{2\pi\tau}$). If the frequency is comparable to or even higher than $\frac{1}{2\pi\tau}$, then the ballistic transport becomes dominant and we cannot apply the "scattering-based" Drude model. Second, the frequency should not be too small, but be large enough to observe the variation in conductivity. This is because the effect of ωτ becomes invisible in complex conductivity $\tilde{\sigma}(\omega) = \frac{\sigma_{dc}}{1-i\omega\tau}$ if ωτ~0. THz frequency can satisfy these two conditions, as probed in early studies[4-6].

In our system, the frequency range of THz-TDS is about 0.5~2.5 THz and the scattering time is about 30 fs. This leads to the value of ωτ as $0.09 < \omega\tau < 0.47$, which satisfies the first condition for applying Drude model. Moreover, as shown in Fig. S3, the variation of complex conductivity can be clearly seen in THz frequency range, which also satisfies the second condition for Drude fitting. Therefore, we can determine the scattering time using the Drude fitting.

**Effect of fitting function and fitting frequency range**

We lastly discuss the accuracy of the Drude fitting in terms of fitting function and fitting frequency range. For Drude fit, we used "instrumental error weighting function" which gives weight to each data inversely proportional to the size of error (that is, data point with smaller error would affect more). The error bars in Figs. 2 and 3 in the main manuscript were the results from the fitting based on the "instrumental error weighting function". We note that the fitting error does not much vary even when we use a "standard equal weighting function", since our complex conductivity data have very small error (the change in $\tau$ depending on the fitting function is about 0.03%). To check the effect of fitting frequency range, we fitted the data by reducing the frequency range to 1 THz < $f$ < 1.5 THz, instead of 0.75 THz < $f$ < 1.75 THz. The resulting $\tau$ does not change within 0.04 %. Considering that the error in $\tau$ is about 0.2 % in our result, the selection of fitting function as well as the fitting frequency range does not much affect the determination of $\tau$.

## Note 4. Confirmation of DC resistivity by four-probe measurement

To check the validity of DC resistivity ($\rho_{dc,THz}$) obtained by THz-TDS, we performed the standard DC resistivity measurement. To this end, we prepared Py wire (1 mm × 4.5 mm) which has same nominal thickness, and performed the four-probe DC resistivity measurement. The measurement was done by using Physical Property Measurement System (PPMS) at various temperatures. The closed symbols in Fig. S4 exhibits the temperature dependent DC resistivity measured by the four-probe method. The obtained DC resistivity is consistent with the $\rho_{dc,THz}$ (open symbols in Fig. S4), confirming the reliability of our THz-TDS measurement. The slight difference between standard four-probe DC measurement and THz-TDS is possibly due to the sample differences

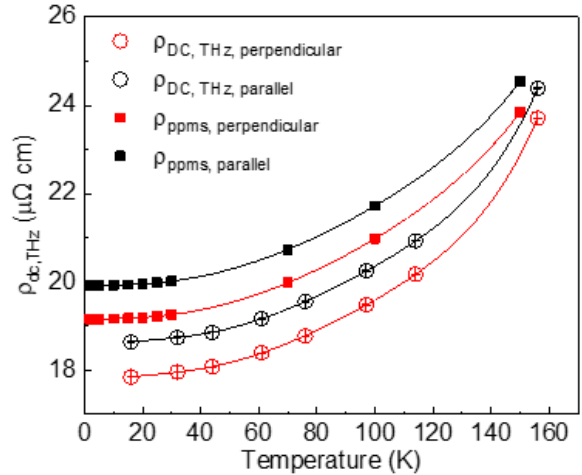

**Fig. S4** Temperature dependent resistivity for parallel and perpendicular geometries of Py (90nm)/Ta (1nm) sample by THz-TDS (open symbols) and standard four-probe DC measurements done by PPMS (closed symbols). Solid lines are standard $T^5$ fitting.

## Note 5. Temperature dependence of *n* and *m\**

To identify the origin of the temperature dependence of $n/m^*$, we have tried the ordinary Hall effect (OHE) measurement and obtained the temperature dependent charge density, $n_{OHE}$. Figure S5a shows that the $n_{OHE}$ increases by about 5% with increasing temperature from $T$ = 16K to $T$ = 155K. The temperature dependence of $m^*$ can be extracted from the measured $n_{OHE}$ [Fig. S5a] and $(n/m^*)_{perp}$ [Fig. 2c in main manuscript]. Here, we used $(n/m^*)_{perp}$ because the perpendicular geometry is consistent with the situation of $n_{OHE}$ measurement. Figure S5b shows that $m^*$ increases by about 13% with increasing temperature from $T = 16K$ to $T = 155K$. These results suggest that both the $n_{OHE}$ and $m^*$, which gives intrinsic contribution to AMR, varies with temperature. We note that the values of $n_{OHE}$ and $m^*$ as well as their temperature dependences are consistent with the previous reports[7-12].

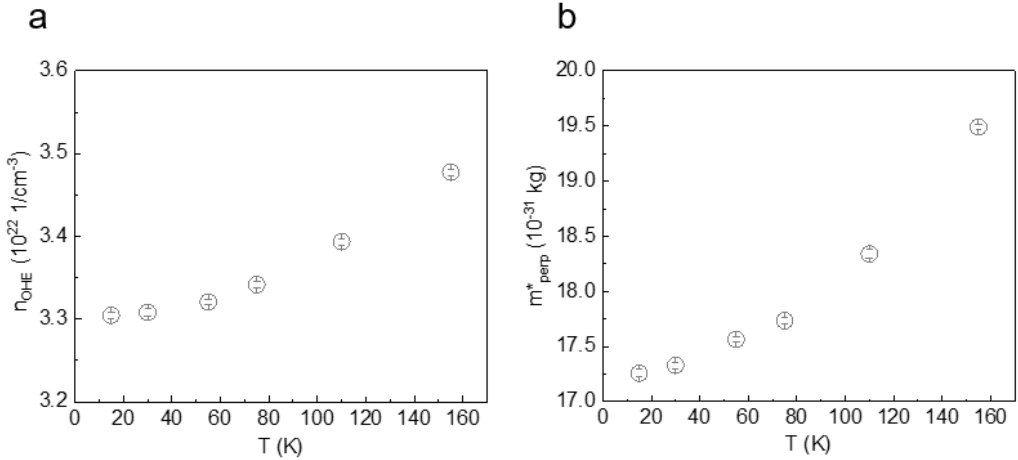

**Fig. S5. a,** Charge concentration, $n_{OHE}$, as a function of temperature obtained from ordinary Hall effect measurement. **b,** effective mass at perpendicular geometry ($m^*_{perp}$) as a function of temperature.

## Note 6. Existence of intrinsic AMR in polycrystalline films

The intrinsic AMR can exist even in the polycrystalline films as follows. In polycrystalline samples, where the crystalline axis is randomly distributed among grains, the dependence of the carrier density $n$ and the effective mass $m^*$ on the wavevector $\mathbf{k}$ and the magnetization direction $\mathbf{m}$ can be phenomenologically written as $n(\mathbf{k}; \mathbf{m}) = n(\mathbf{k} \cdot \mathbf{m})$ and $m^*(\mathbf{k}; \mathbf{m}) = m^*(\mathbf{k} \cdot \mathbf{m})$, where $\mathbf{k}$ and $\mathbf{m}$ appear only through their inner product (since there is no preferential direction due to the polycrystalline nature). If the inversion symmetry is not broken macroscopically in polycrystalline samples, we have more stringent functional dependence: $n(\mathbf{k}; \mathbf{m}) = n((\mathbf{k} \cdot \mathbf{m})^2)$ and $m^*(\mathbf{k}; \mathbf{m}) = m^*((\mathbf{k} \cdot \mathbf{m})^2)$, which gives rise to the anisotropy that is discussed in our manuscript. These expressions meet all the symmetry properties required by polycrystalline ferromagnetic samples. Therefore, a certain amount of anisotropy in $n/m^*$ is expected to occur even in polycrystalline ferromagnets.

Although the anisotropy in $n/m^*$ would be larger in single-crystal samples than in polycrystalline samples, as argued above, even in polycrystalline films, the finite anisotropy is expected to exist in $n/m^*$, meaning that the change of the band structure due to the external field via spin-orbit coupling gives rise to a finite anisotropy in $n/m^*$ even after being averaged among grains with different crystal orientations. We note that similar argument has been employed to show the existence of the intrinsic anomalous Hall effect in polycrystalline films, where the angular average of intrinsic contributions for all crystalline directions was found to be finite in good agreement with the measured intrinsic contribution for polycrystalline films[13].

In polycrystalline metals, there are two important length scales, the electron mean free path and the average grain size, within which the electronic band structure is well defined locally. It is reasonable to classify the contributions to AMR from the anisotropy of $n/m^*$ as intrinsic ones if the grain size is larger than the mean free path so that electron dynamics can manifest the local band structure. In other words, as long as the electrons feel the band structure of each grain, we expect that there is a finite (average) intrinsic contribution to AMR.

To further support our claim, we have checked the microstructure of our Py film by X-ray diffraction (XRD) measurement. Figure S6 shows the XRD spectra for our polycrystalline Py film. A peak was clearly observed at $2\theta \sim 44°$ in the spectra, indicating the fcc (111) crystalline phase of Py[14]. The grain size can be estimated from the spectra based on the Scherrer formula, and we found that the average grain size was about 12.8 nm. As the grain size is larger than the electron mean free path of about 6 nm in Py, one can expect that the electrons feel the band structure of each grain, which could give rise to the average intrinsic contribution (Here, we estimated the electron mean free path from the Fermi velocity[15] ($v_F = 0.22 \times 10^6 \, m/s$) and measured scattering time ($\tau = 23 \sim 30 \, fs$ in Fig. 2(b) in our manuscript)).

As for the microscopic origin of intrinsic contribution, we think that it originates from the magnetization dependent *d*-orbital anisotropy which affects electron conduction through *spd*- hybridization. A quantitative calculation, however, requires a more rigorous transport theory including *spd*-hybridization, which is beyond the scope of this work.

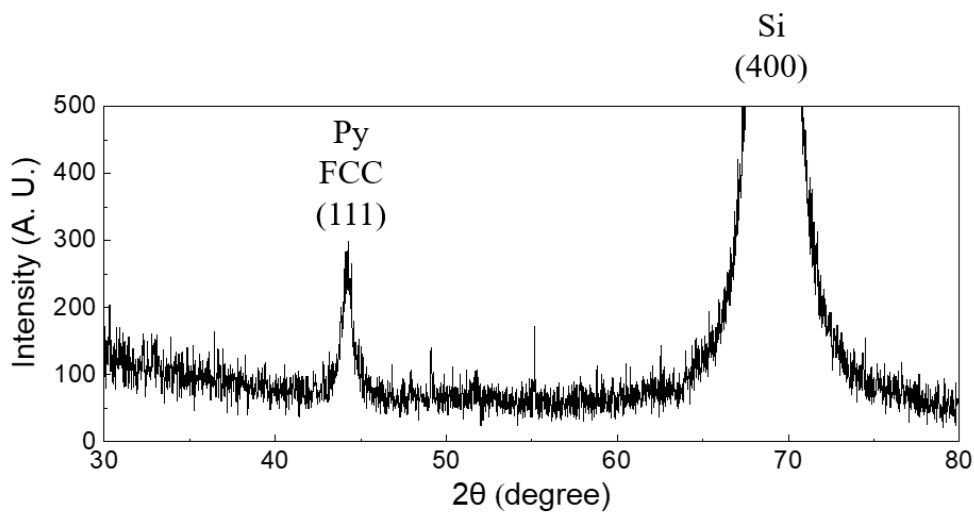

**Fig. S6.** The XRD θ − 2θ scan spectra of our permalloy sample. A peak at 2θ~44° indicates the fcc (111) crystalline phase of Py. A large peak at 2θ~70° arises from the substrate Si (400) [16].